\documentclass[
    ,final            % use final for the camera ready runs
  ]
  {aipproc}

\layoutstyle{8x11single}

\begin{document}

\title{Lattice QCD and the Balkan physicists contribution}


\author{Artan Bori\c{c}i}{
  address={Faculty of Natural Sciences, University of Tirana, Bvd. Zog I, Tirana, Albania}
}

%\author{<author2>}{
%  address={<common address for author2 and author3>}
%}

%\author{<author3>}{
%  address={<common address for author2 and author3>}
%  ,altaddress={<author1 address>} % additional visiting address
%}

\begin{abstract}
This is a paper based on the invited talk the author gave at the 9th Balkan Physical Union conference. It contains some of the main achievements of lattice QCD simulations followed by a list of Balkan physicists who have contributed to the project.
\end{abstract}

\maketitle

%%%%%%%%%%%%%%%%%%%%%%%%%%%%%%%%%%%%%%%%%%%%
%% MAINMATTER
%%%%%%%%%%%%%%%%%%%%%%%%%%%%%%%%%%%%%%%%%%%%

\subsection{Prehistory of lattice QCD}

Lattice Quantum Chromodynamics (QCD) is in its 41st year since the seminal paper of Kenneth Wilson in 1974, {\it Confinement of Quarks} \cite{Wilson1974}.\footnote{For a historical review of the lattice theory see the paper of Creutz \cite{Creutz2004}. For a more recent review see the one written by Ukawa \cite{Ukawa2015}.} From that time remarkable events took place in our understanding of strong interactions. The lattice definition of QCD is an important milestone in the development of particle physics theories. Thanks to the asympototic freedom and the reflection positivity lattice QCD is {\it the} QCD as opposed to the continuum formulation which is not well-defined mathematically beyond perturbation theory.

Historically, it was the quark model that paved the way for the modern theory of the strong interactions. In 1963 Gell-Mann and Zweig proposed that the low lying hadron masses may be described as many body representations of the $SU(2)$ and $SU(3)$ flavor groups. Later, Greenberg, Han and Nambu introduced the color degree of freedom postulating that each quark flavor transforms under the fundamental representation of the $SU(3)$ color group and that all hadrons are color singlets. The deep inelastic lepton-nucleon scattering experiments at SLAC that took place in the end of 1960s revealed that quarks or partons, as Feynman called them, behave like free particles at high energies. It was then clear that any quantum field theory of strong interactions should have this property. Although non-Abelian gauge theories were known since 1954 their renormalizibility was shown only at the beginning of 1970s. This development led Gross and Wilczek as well as Politzer in 1973 to show that non-Abelian gauge theories are asymtotically free. Quarks are however confined into hadrons and this property was not shown until Wilson proved it for the strongly coupled lattice gauge theories.

\subsection{The lattice}

The Wilson formulation allows a fully non-perturbative treatment of lattice gauge theories. Yet, the expectation value of a typical observable of the theory is not known analytically. Lattice QCD in not an exception, eg. the three-dimensional Ising model and many more models in statistical mechanics and field theory have no analytical solutions. The only possibility that remains is to use Markov Chain Monte Carlo algorithms like the local heat bath algorithm that was devised by Creutz in 1979 \cite{Creutz1980}. In that calculation he was able to show that the string tension of the $SU(2)$ and $SU(3)$ lattice theories scales as expected by the asymptotic freedom demonstrating thus in principle that lattice gauge theories can be investigated by means of Monte Carlo simulations. A series of proposals for the calculation of the glueball as well as hadron masses followed (see Rebbi's book for more details \cite{Rebbi1983}).

\subsection{Moore's law}

\begin{figure}\label{alpha_spectrum}
\includegraphics[height=.25\textheight]{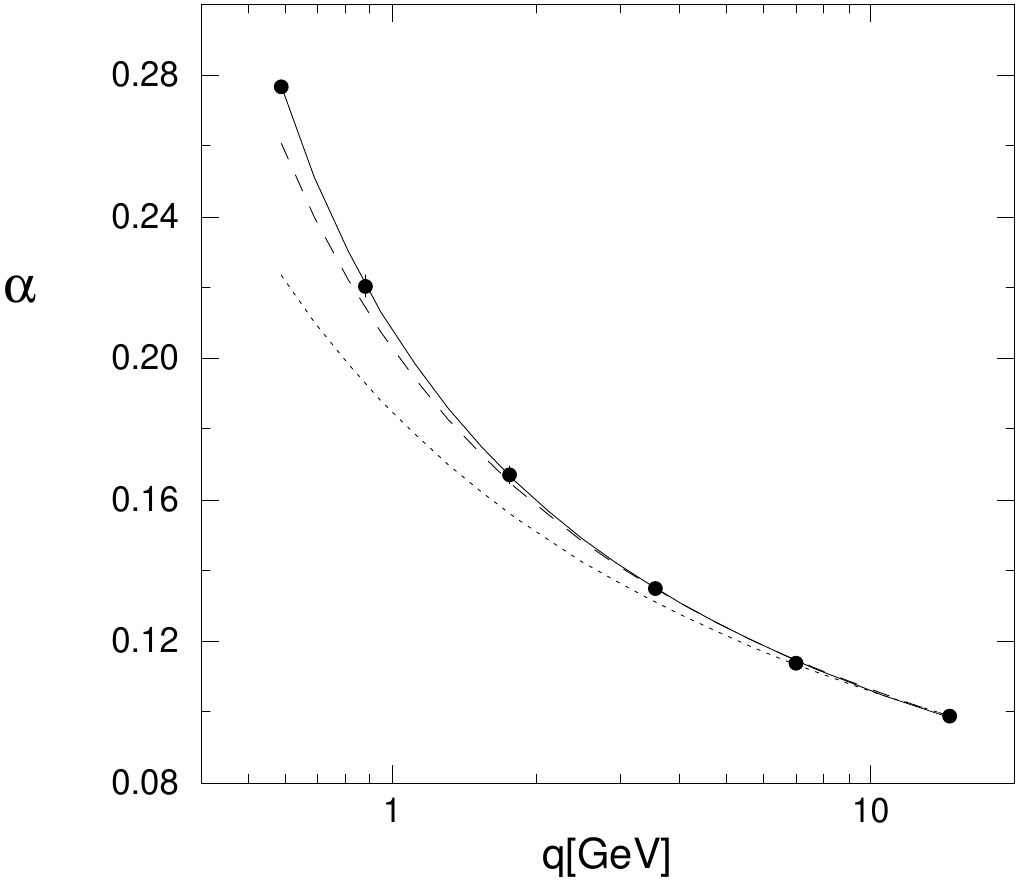}~~~~~~~~~~~~~~\includegraphics[height=.25\textheight]{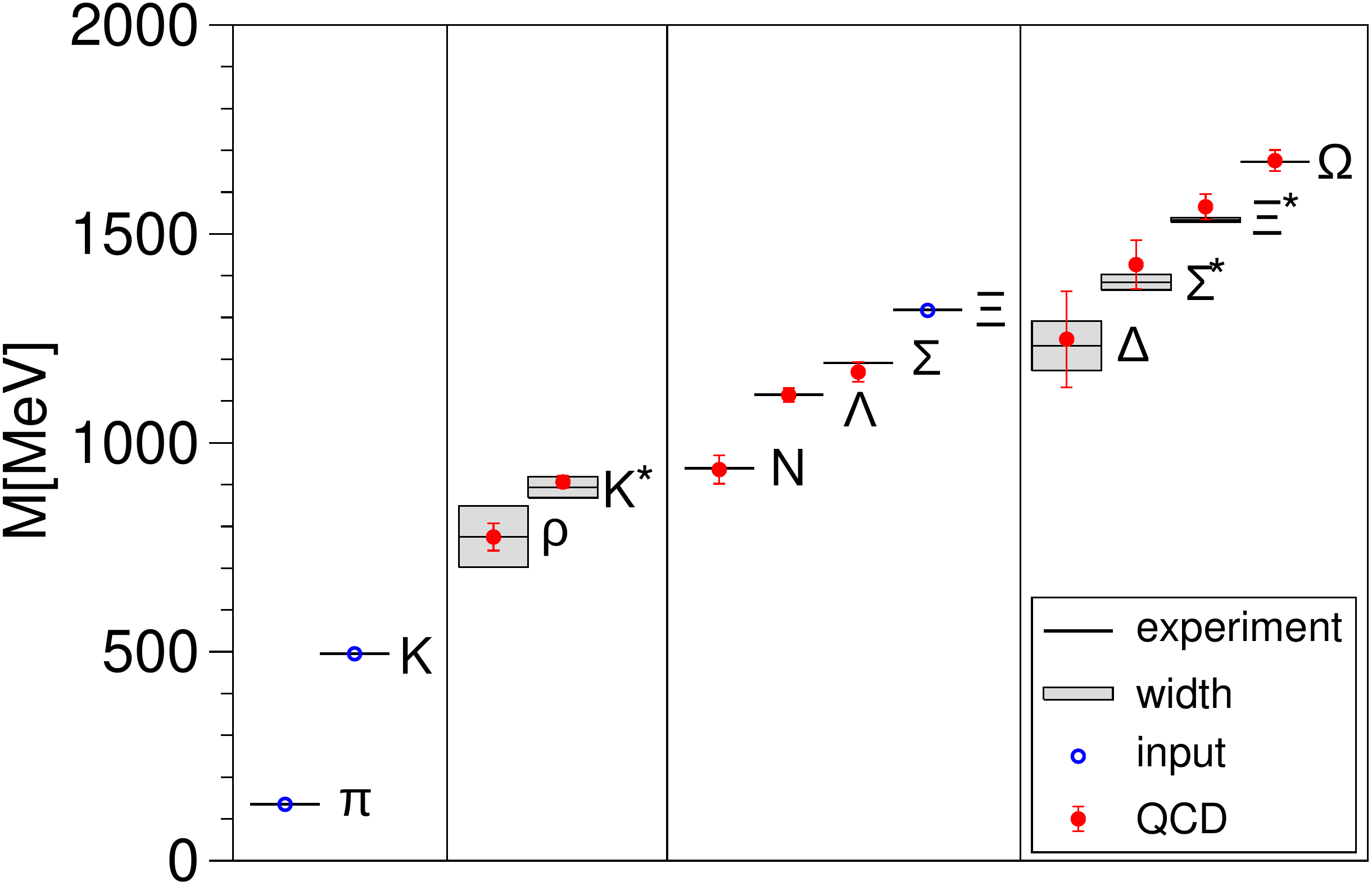}
\caption{The running coupling constant of the $SU(3)$ theory together with the one-loop and two-loop results (left plot) \cite{Luscheretal1994} and the light hadron spectrum of QCD (right plot) \cite{BMW2008}.}
\end{figure}

The first calculation of Creutz required around $1$ MFlops to compute the plaquette of the $SU(3)$ theory on a $6^4$ lattice, which could run on the VAX-11/780. However, the long distance effects of QCD require $30^4$ lattices at a $0.1$ fm lattice spacing, which means a GFlops machine. The quark-antiquark potential at those distances was calculated 12 years later by Bali and Schilling on the CM-2 machine \cite{BaliShilling1993}, whereas the running coupling of the $SU(3)$ theory was calculated one year later by the Alpha Collaboration \cite{Luscheretal1994}. The calculation of the hadron spectrum without quark loops took $11$ GFlops-years to $300$ GFlops-years to complete on the GF11 machine in 1993 and on the CP-PACS machine in 1999 respectively \cite{QuenchedSpectrum}. The full QCD spectrum was calculated in 2008 on the Blue Gene/P machine with the performance of around 100 TFlops \cite{BMW2008}. Some results of these calculations are shown in Figure \ref{alpha_spectrum}. It has taken about 30 years from the first $SU(3)$ calculation till the full QCD one. It is Moore's law, the exponential increase of computing power with time, that enabled lattice QCD simulations to reach this milestone.

%%%%%%%%%%%%%%%%%%%%%%%%%%%%%%%%%%%%%%%%%%%%
%% SAMPLE TABLE
%%
%% Shows the use of \tablehead and \tablenote
%% macros
%%%%%%%%%%%%%%%%%%%%%%%%%%%%%%%%%%%%%%%%%%%%

%\begin{table}
%\begin{tabular}{lrrrr}
%\hline
%  & \tablehead{1}{r}{b}{Single\\outlet}
%  & \tablehead{1}{r}{b}{Small\tablenote{2-9 retail outlets}\\multiple}
%  & \tablehead{1}{r}{b}{Large\\multiple}
%  & \tablehead{1}{r}{b}{Total}   \\
%\hline
%1982 & 98 & 129 & 620    & 847\\
%1987 & 138 & 176 & 1000  & 1314\\
%1991 & 173 & 248 & 1230  & 1651\\
%1998\tablenote{predicted} & 200 & 300 & 1500  & 2000\\
%\hline
%\end{tabular}
%\caption{Average turnover per shop: by type
%  of retail organisation}
%\label{tab:a}
%\end{table}

\subsection{Numerical algorithms}

QCD has not been solved yet. It will take some time until precise calculations of a large number of strongly interacting particle properties become pervasive. However, based on the great progress witnessed in the development of numerical algorithms, it is not expected to take another 30 years for the lattice QCD to achieve this goal. Looking back in time, the Interdisciplinary Project Center for Supercomputing (IPS) of the ETH Z\"urich was the ideal environment for a lattice group to develop numerical algorithms. Led by de Forcrand and influenced by a new version of the BiCGStab algorithm of Gutknecht, the Z\"urich group together with the Wuppertal group, led by Schilling in collaboration with Frommer, made a pioneering work on the non-hermitian solvers for lattice QCD, a research which resulted in large savings of computer time for quark propagator calculations \cite{BiCGStab1994}. From that time numerical algorithms for lattice QCD have attracted a pleiad of mathematicians including Gutknecht, Frommer, Golub, Saad, van der Vorst, Higham, Manteuffel and many more in a series of workshops of Numerical Analysis and Lattice QCD \cite{QCDNA2000&2005}. Among the recent achievements of this collaboration is the adaptive multigrid algorithm for the inversion of the Wilson operator \cite{Adaptive_MG}. When these ideas appeared, L\"uscher had already developed the concept of local coherence of the low lying modes of the Dirac operator resulting in a deflation based algorithm with large savings in computer time \cite{LocalCoherence}. Figure \ref{local_coherence} shows clearly that these algorithms accelerate inversions by a very large factor.

\subsection{Simulation algorithms}

\begin{figure}\label{local_coherence}
\includegraphics[height=.29\textheight]{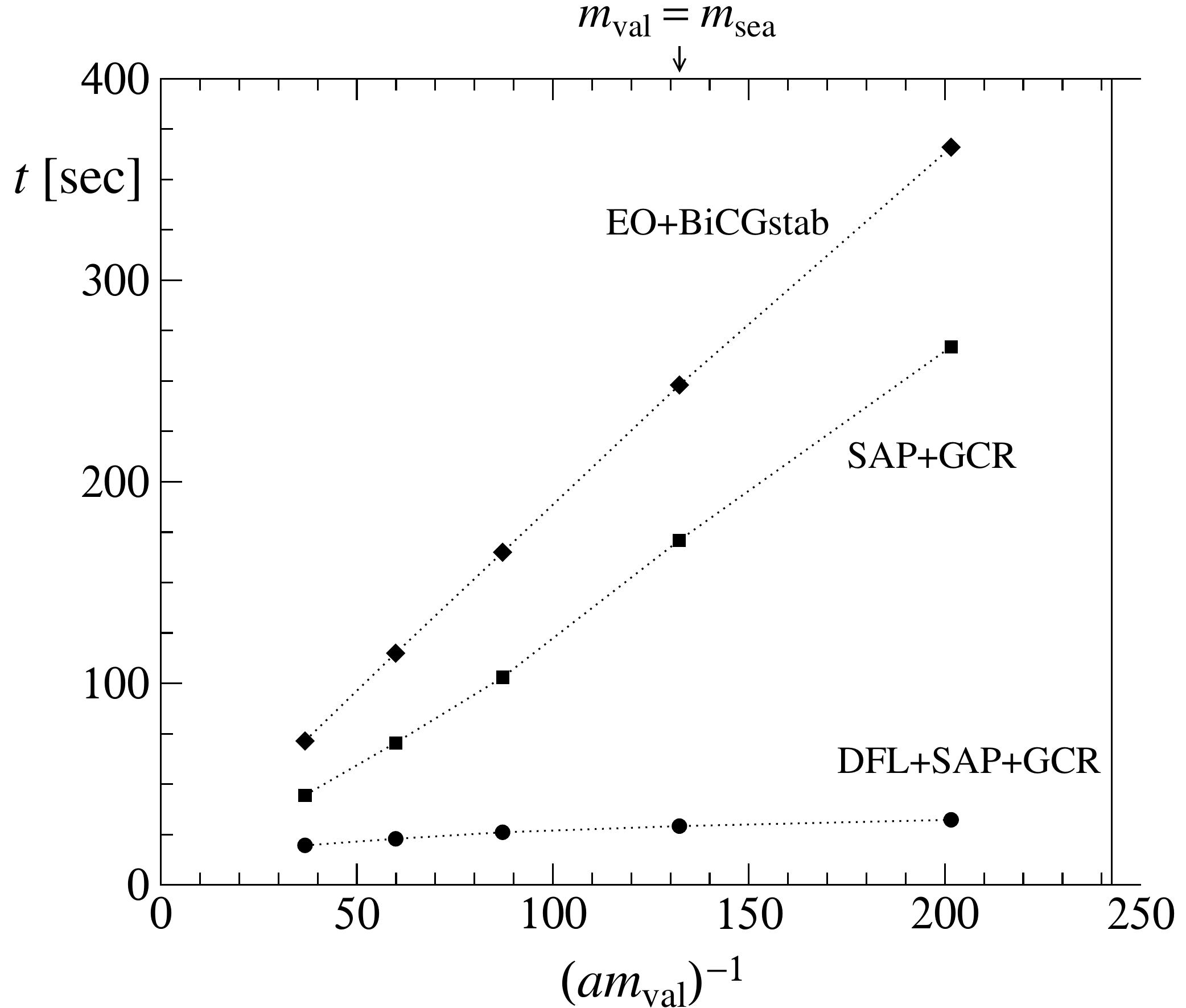}\includegraphics[height=.29\textheight]{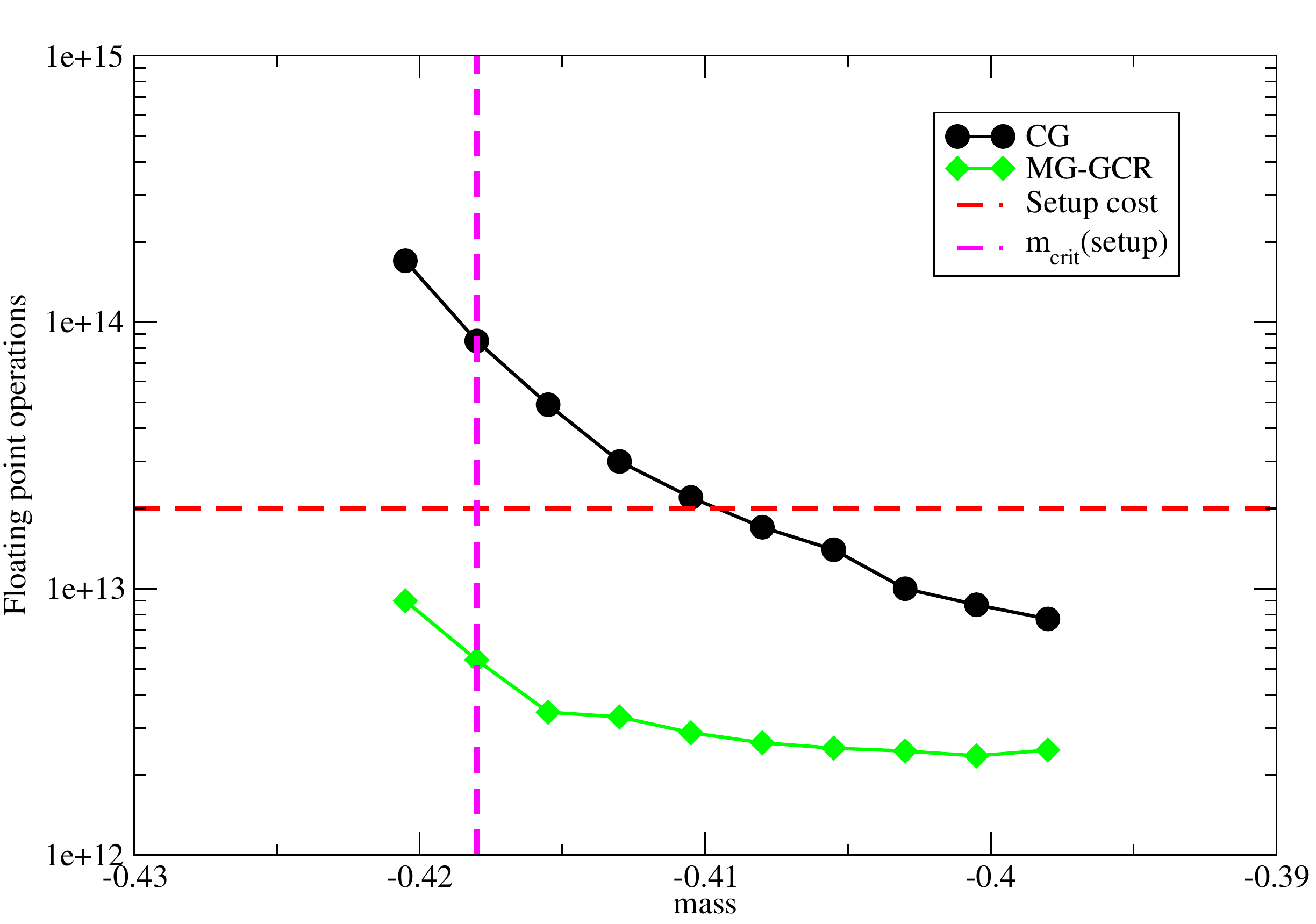}
\caption{The local coherence of low lying modes of the Dirac operator is the basic property that allows an efficient deflation of the Dirac operator (left plot) \cite{LocalCoherence}. Essentially the same algorithm may be constructed using the ideas of adaptive multigrid solvers (right plot, Babich et. al.) \cite{Adaptive_MG}. Both plots show a gain in computer time by more than an order of magnitude.}
\end{figure}

Simulation of QCD with dynamical quarks has always been challenging. The Dirac operator must be inverted a large number of times along a trajectory generated by the Hybrid Monte Carlo algorithm \cite{HMC1987}. The amount of computer resources using this algorithm at the physical point is of the order of Petaflop-years. The situation remained this way until L\"uscher made two important improvements: using a domain decomposition technique he could accelerate inversions by a factor of five (see SAP+GCR performance in Figure \ref{local_coherence}) which translates to an accelerated Hybrid Monte Carlo algorithm by the same order of magnitude \cite{Luscher_SAP}. Deflation of the low modes of the Dirac operator \cite{LocalCoherence} allowed him to devise the deflation accelerated simulation algorithm which gives an independent improvement by a factor five or so \cite{Luscher_defl_hmc}. While the exponential increase in computing power for one invested monetary unit remains a very important driving force in lattice QCD, the algorithmic advances have been at least as important, particulary so in the last 15 years \cite{Luscher_Les_Houches2009}.

\subsection{Chiral symmetry and the fifth dimension}

Chiral symmetry is an important symmetry of strong interactions. However, on the lattice, a Weyl fermion may not exist without its opposite chirality partner. This means that a Dirac fermion may not exist without its doubler. Wilson gave the doublers a mass proportional to the lattice cutoff destroying thus the normality of the Wilson-Dirac operator:
$$
[D_W,D_W^*]\ne0\ .
$$
This way, one has to fine tune the bare quark mass in order to restore the chiral symmetry. It is easy to see that any $\gamma_5$-hermitian lattice Dirac operator is normal if it satisfies the Ginsparg-Wilson relation \cite{GW1982}:
$$
\gamma_5D+D\gamma_5=a\;D\gamma_5D\ ,
$$
where $a$ is the lattice spacing. This simple relation has apparently only one formal solution, the Overlap Dirac operator of Neuberger \cite{Neuberger1998}:
$$
a\;D=I+D_W(D_W^*D_W)^{-1/2}.
$$
Prior to this discovery a five dimensional theory of chiral fermions, the Domain Wall fermions, was formulated \cite{DWF}. Since the Overlap operator is the Dirac operator of the four dimensional effective theory of Domain Wall fermions at any finite lattice spacing \cite{Borici1999}, lattice QCD may be regarded as an effective theory of a higher dimensional theory. Five dimensional theories are even more expensive to simulate and the hope is that various multigrid ideas will work in this case as well \cite{MG_Overlap}. Another approach is to simulate local theories with the minimum number of doublers at the expense of a broken hypercubic symmetry \cite{MDF}. Since the restoration of the broken symmetry involves the fine tuning of certain lattice operators \cite{Capitani_etal2010} new versions of such fermions with manifest unitarity should be sought \cite{Borici2014}.

%, the multigrid algorithm for the overlap operator \cite{MG_Overlap}

%\paragraph{<A subsubsubsection>}

\subsection{Balkan physicists contribution in lattice QCD}

Lattice physicists born in the Balkans and its close neighborhood include:
I.O. Stamatescu (Heidelberg): {\it RG flow, Topology, finite density and finite temperature}.
H. Neuberger (Rutgers): {\it Chiral fermions, computation of the overlap}.
Y. Shamir (Tel Aviv): {\it Domain wall fermions, beyond SM}.
B. Svetitsky (Tel Aviv): {\it Finite temperature, beyond SM}.
E.T. Tomboulis (UCLA): {\it Topology, confinement}.
T. \c{C}elik (Ankara): {\it deconfinement, percolation}.
A. Vladikas (Rome): {\it Non-perturbative renormalisation, heavy-meson decays}.
P. Dimopoulos (Rome): {\it Spectroscopy, pseudoscalar decay constants}.
C. Alexandrou (Nicosia): {\it Heavy-light quarks, B-mesons, light baryons}.
H. Panagopoulos (Nicosia): {\it Topology, finite temperature, perturbation theory}.
G. Koutsou (Nicosia): {Spectroscopy, form factors}.
M. Constantinou (Nicosia): {Form factors, non-perturbative renormalization}.
A. Tsapalis (Athens): {\it Form factors, multi-quark potentials}.
K. Anagnostopoulos (Athens): {\it Supersymmetric QM, large N super Yang-Mills}.
K. Farakos (Athens): {\it Elektroweak pahse, gravity}.
P. Vranas (Livermore): {\it finite temperature, domain wall fermions}.
K. Orginos (Williamsburg): {\it Spectroscopy, improvements, algorithms}.
D. Becirevic (Orsay): {\it Spectroscopy, form factors, BB mixings}.
L. Levkova (Utah): {\it Spectroscopy, decay constants, finite temperature}.
A. Alexandru (Washington): {\it Spectroscopy, finite density, topology}.
S. Prelovsek (Ljubljana): {\it Spectroscopy, scattering}.
M. \"Unsal (Raleigh): {\it Topology, sypersymmetry, QCD-like theories}.
I. Hip (Zagreb): {\it Ginsparg-Wilson fermions, topology}.
M. Marinkovic (CERN): {\it Quark masses, topology, finite density}.
D. Xhako (Tirana): {\it Domain wall fermion algorithms}.
D. Djukanovic (Mainz): {\it Spectroscopy, chiral perturbation theory, QED}.
R. Zeqirllari (Tirana): {\it Minimally doubled fermions}.
T. Sulejmanpasic (Raleigh): {\it Topology}.
L. Leskovec (Ljubljana) : {\it Spectroscopy}.
M. Vidmar (Ljubljana): {\it Spectroscopy}.
S. Zafeiropoulos (Frankfurt): {\it Dirac spectrum}.
N. Bozovic (Wuppertal): {\it Inversion algorithms}.
M. Blazenka (Zagreb): {\it Decay constants}.
K. Hadjiyiannakou (Nicosia): {Spectroscopy}.
A. Bori\c{c}i (Tirana): {\it Numerical algorithms, Overlap/Domain wall fermions, minimally doubled fermions}.

\bigskip

The author would like to thank M. Creutz, Ph. de Forcrand and M. L\"uscher for their invaluable comments on this contribution. He is indebted to M. Marinkovi\'{c}, M. M\"uller-Preusker, B. Svetitsky and A. Vladikas for sharing their information on lattice physicists from the Balkans.

%%%%%%%%%%%%%%%%%%%%%%%%%%%%%%%%%%%%%%%%%%%%%%%%
%% BACKMATTER
%%%%%%%%%%%%%%%%%%%%%%%%%%%%%%%%%%%%%%%%%%%%%%%%

%\begin{theacknowledgments}
%\end{theacknowledgments}


\bibliographystyle{aipproc}   % if natbib is available

\begin{thebibliography}{9}

\bibitem{Wilson1974}

K. G. Wilson, Phys. Rev. D10 (1974) 2445.

\bibitem{Creutz2004}

M. Creutz, in {\it 50 Years of Yang-Mills Theory}, G. 't Hooft (edt.) (2005) 357. 

\bibitem{Ukawa2015}

A. Ukawa, {\it Kenneth Wilson and lattice QCD}, arXiv:1501.04215.

\bibitem{Creutz1980}
M. Creutz, Phys. Rev.  D21 (1980) 2308; Phys.Rev.Lett. 45 (1980) 313.

\bibitem{Rebbi1983}

C. Rebbi, Lattice Gauge Theories And Monte Carlo Simulations, World Scientific 1983.

\bibitem{BaliShilling1993}

G. S. Bali, K. Schilling, Phys. Rev. D47 (1993) 661.

\bibitem{Luscheretal1994}

M. L\"uscher et. al., Nucl. Phys. B413 (1994) 481.

\bibitem{QuenchedSpectrum}

F. Butler et.al., Phys. Rev. Lett. 70 (1993) 2849; S. Aoki et. al., Phys. Rev. Lett. 84 (2000) 238.

\bibitem{BMW2008}

S. Durr et. al., Science 322 (2008) 1224.

\bibitem{BiCGStab1994}

A. Bori\c{c}i, Ph. de Forcrand, in Physics Computing '94 (1994) 711; A. Frommer et. al., Int.J.Mod.Phys. C5 (1994) 1073.

\bibitem{QCDNA2000&2005}

A. Frommer et. al. (edts), {\it Numerical Challenges in Lattice Quantum Chromodynamics}, Springer 2000; A. Bori\c{c}i et. al. (edts), {\it QCD and Numerical Analysis III}, Springer 2005.

Comput.Phys.Commun. 146 (2002) 203

\bibitem{Adaptive_MG}

R. Babich et. al., Phys.Rev.Lett. 105 (2010) 201602; A. Frommer et. al., arXiv:1307.6101.

\bibitem{LocalCoherence}

M. L\"uscher, JHEP0712:011,2007.

\bibitem{HMC1987}

S. Duane, A. D. Kennedy,  B. J. Pendleton, D. Roweth, Phys. Lett. B195 (1987) 216.

\bibitem{Luscher_SAP}

M. L\"uscher, Comput.Phys.Commun. 156 (2004) 209-220; Comput.Phys.Commun. 165:199-220,2005.

\bibitem{Luscher_defl_hmc}

M. L\"uscher, JHEP 0712:011,2007.

\bibitem{Luscher_Les_Houches2009}

M. L\"uscher in {\it Modern perspectives in lattice QCD}, L. Lellouch et. al. (edts.), Oxford (2011) 331.

\bibitem{GW1982}

P. H. Ginsparg and K. G. Wilson, Phys. Rev. D25 (1982) 2649.

\bibitem{Neuberger1998}

H. Neuberger, Phys.Lett. B417 (1998) 141.

\bibitem{DWF}

D.B. Kaplan, Phys.Lett. B288 (1992) 342-347; V. Furman and Y. Shamir, Nucl.Phys. B439 (1995) 54.

\bibitem{Borici1999}

A. Bori\c{c}i, Nucl. Phys. B(Proc. Suppl.) 83 (2000) 771; A. Bori\c{c}i, in {\it Lattice Fermions and Structure of the Vacuum}, V. Mitrjushkin and G. Schierholz (edts.), Kluwer Academic Publishers, 2000.

\bibitem{MG_Overlap}

A. Bori\c{c}i, Phys.Rev. D62 (2000) 017505; D. Xhako and A. Bori\c{c}i, Am. J.Phys.Appl. 2 (2014) 67; J. Brannick et. al., arXiv:1410.7170.

\bibitem{MDF}

L. H. Karsten, Phys. Lett. B 104, 315 (1981); F. Wilczek, Phys.Rev.Lett.59:2397,1987; M. Creutz, JHEP 0804, 017 (2008); A. Bori\c{c}i, Phys.Rev.D78:074504,2008.

\bibitem{Capitani_etal2010}

S. Capitani et. al., JHEP 1009 (2010) 027.

\bibitem{Borici2014}

A. Bori\c{c}i, PoS LATTICE2014 (2014) 332.

\end{thebibliography}

%%%%%%%%%%%%%%%%%%%%%%%%%%%%%%%%%%%%%%%%%%%
%% The following lines show an example how to produce a bibliography
%% without the help of the BibTeX program. This could be used instead
%% of the above.
%%%%%%%%%%%%%%%%%%%%%%%%%%%%%%%%%%%%%%%%%%%

\end{document}